**On Ambient-light-induced intermolecular Coulombic decay in unbound pyridine monomers**

Shaivi Kesari, Amol Tagad and G. Naresh Patwari*

Department of Chemistry, Indian Institute of Technology Bombay, Mumbai 400076 INDIA

Email: naresh@chem.iitb.ac.in

In a recent report, Barik *et al.*[1] proposed that single-photon excitation of pyridine molecules with 266 nm (4.66 eV) radiation, results in the ionization of pyridine monomer ($IP_0$ = 9.24 eV) following an association of three pyridine monomers in the excited state, which were otherwise unbound in the neutral ground state. The observation of pyridine monomer cation along with an isotropic emission of low kinetic energy electrons (up to 0.6 eV) could only be attributed to be arising out of intermolecular Coulombic decay (ICD)[2] of unbound pyridine monomers. However, the proposed model raises several reservations regarding the probability of such molecular association, especially in the excited state, considering the collision-free conditions which is inherent to free-jet expansion,[3] low absorption cross-section ($19.3 \times 10^{-19}$ cm$^2$) of pyridine at 266 nm,[4] and a short lifetime of about 40-50 ps of the excited state.[5–8] Further, the $S_2$ to $S_0$ non-radiative relaxation rate constant is about $11.4 \times 10^{11}$ s$^{-1}$ (corresponding lifetime is about 0.9 ps) and the electronic relaxation from the $S_1$ to $S_0$ state is dominant and suppresses the intramolecular vibrational redistribution (IVR) in the $S_1$ state.[8] However, a much longer lifetime of 0.5 ns was considered, the origin of which cannot be traced. Additionally, it was assumed that the system could then transit back and forth between $S_1$ and $S_0$, hence we do have a longer lifetime component as well. The back-and-forth population transfer between the $S_1$ and $S_0$ states with energies in excess of 30,000 cm$^{-1}$ above the $S_0$ minimum, wherein the IVR is dissipative, has not been experimentally demonstrated. On the contrary, the appearance of the quantum beats in the $S_1$ state solely within the restrictive IVR regime with excess energies of around 2500 cm$^{-1}$ has been established.[9,10]

Barik *et al.*[1] evaluated the number density of the excited state pyridine molecules in the interaction volume of 0.0981 cm$^3$ to be about $2.15 \times 10^{11}$ per pulse, which suggests that the probability of finding two and three independently excited pyridine molecules in a sphere of 10 Å radius is about $9.0 \times 10^{-9}$ and $8.1 \times 10^{-17}$, respectively. The total energy in two excited pyridine molecules (9.32 eV) is higher than the ionization potential of the pyridine monomer (9.24 eV), and the appearance energy of the pyridine monomer cation from the dimer is 9.25 eV,[11] which could explain the absence of the pyridine dimer cation as noted by Barik *et al.*[1] However, the pyridine monomer cation signal arising out of a two-body association will outweigh any mass signals that arise out of a three-body association since the probability of a two-body association is several orders of magnitude higher than a three-body association. It appears that the three-photon dependence of the pyridine cation signal and a maximum electron kinetic energy of 0.6 eV has misled the authors to ignore the two-body association.[1] Further, the three-photon dependence of the pyridine cation signal was interpreted to be originating from single-photon absorption by three independent pyridine molecules. However, independent single-photon absorption by three pyridine molecules must show single-photon dependence of signal and not three-photon dependence as reported.[1] Interestingly, a similar report by Barik *et al.*[12] wherein a five-body association consisting of four excited pyridine molecules and an argon atom leading to the formation of argon ion shows single-photon dependence. Thus, the signals of pyridine cation from the three-body association and of argon cation from the five-body association show markedly different dependence on the number of photons, and therefore the proposed model is inconsistent.[1,12] Additionally, a three-centre ICD process must lead to the formation of a pyridine dimer cation and not a pyridine monomer cation, since the ionization energy of the dimer is lower than the monomer, and the monomer carries the excess energy.[13]

Addressing the question "How isolated are molecules in a molecular beam?" Lubman *et al.*[3] noted that molecules in a free-jet expansion can be considered isolated in most practical cases with times scales of up to several microseconds with collision cross-sections of less than 10 Å$^2$. Therefore, based on the short excited state lifetimes (about 40-50 ps), the low number density of the excited pyridine monomers and persistently long collision-less conditions of the free-jet expansion, the observed mass signals and their dependence on the number of photons, and thus "ambient-light-induced intermolecular Coulombic decay in unbound pyridine monomers" is a misrepresentation on several counts. Additionally, the phenomenon that leads to the ionization of pyridine monomer by an association of three excited pyridine molecules, which were otherwise unbound in the neutral ground state, should be characterized as a special case of penning ionization, rather than intermolecular Coulombic decay.

The ambiguity in interpreting the experimental results by Barik *et al.*[1] and concurrent misleading conclusions has prodded further investigations by carrying out a complementary set of experiments, wherein the translational energy of the pyridine cations was measured using velocity map ion imaging technique following 266 nm excitation, and the results are presented in Figure 1. The mass spectra (Figure 1 and Figure S1) consistently show the presence of pyridine monomers, dimers and even higher clusters in the presence of helium buffer gas. Measurements in the presence of helium buffer gas yield a broad fragment translational energy distribution for the pyridine monomer cation with the laser fluence of about $10^5$ W $cm^{-2}$, which is indicative of a photo-dissociation process and exhibits three-photon dependence. On the other hand, neat expansion of the pyridine gas does not yield signals of dimer and higher clusters, while the pyridine cation signal shows two-photon dependence with the laser fluence around $10^5$ W $cm^{-2}$. The corresponding near-zero translational energy distribution profile confirms the resonantly enhanced multiphoton ionization (REMPI) process, which was deemed non-operational by Barik *et al.*[1] However, the most interesting observation is that the pyridine cation signal shows 2.3 photon dependence at an elevated laser fluence of around $10^6$ W $cm^{-2}$ in the presence of helium buffer gas and the corresponding translational energy distribution for the pyridine monomer cation exhibits characteristics that combine both REMPI and photo-dissociation processes. Contrasting the mass spectra shown in Figure 1, the mass spectra reported by Barik *et al.*[1] showed several signals between 79 (pyridine monomer) and 158 amu (pyridine dimer), which is questionable based on the three-body ICD, since pyridine dimer cation is either non-covalently bound[14] or hemi-bonded,[11] wherein the dissociation threshold for the neutral pyridine dimer to yield pyridine monomer cation is 9.25 eV.[11] Therefore, primary fragmentation of pyridine dimer cation must yield pyridine monomer cation and secondary fragmentation yields mass signals lower than 79 amu. Moreover, the reported absence of intact dimer cation does not preclude the fact that "absence of evidence cannot be the evidence for absence". On the other hand, various signals in the mass spectra reported by Barik *et al.*[1] originate from ion-molecule collisions in the extraction zone between neutral pyridine molecules and the pyridine monomer cation, as illustrated by the results depicted in Figure 2, wherein a noticeable increase in the signals at 91, 106 and 110 amu was observed with varying the electrostatic fields in the extraction zone. Ion-molecule reactions are in general known to have much larger crossections[15] and produce fragments with molecular weight larger than the parent molecules.[16,17] Further, several mass fragments below 79 amu reported by Barik *et al.*[1] are not part of the standard mass spectrum of pyridine,[18] and could originate as a consequence of ion-molecule reactions.

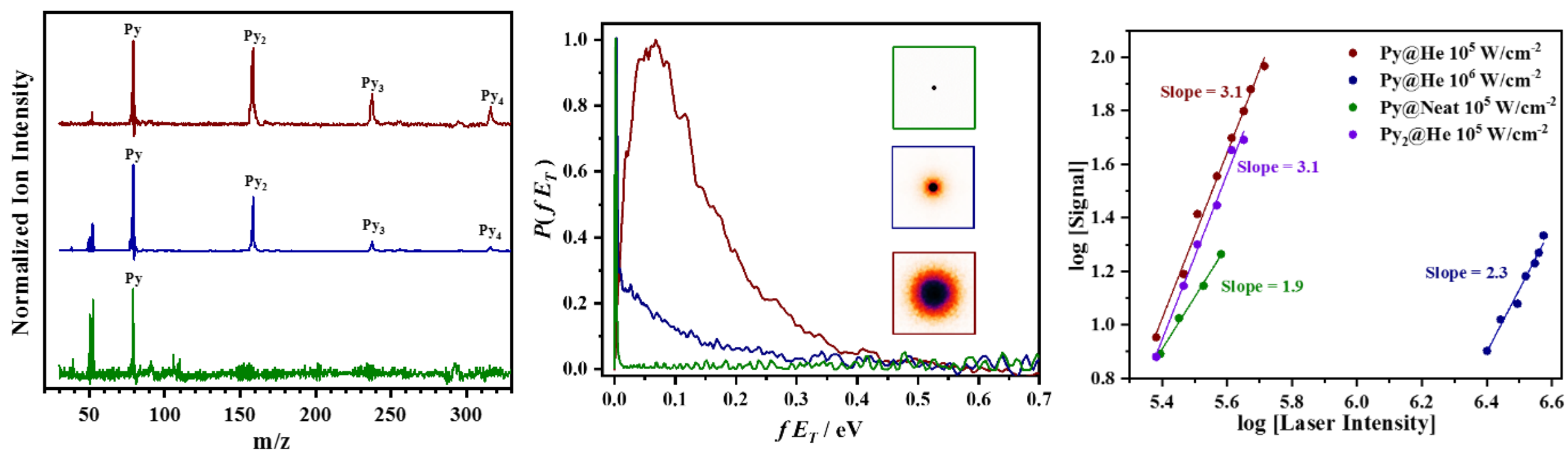

Figure 1. (Left panel) Mass spectrum of pyridine recorded under various conditions. The top (brown) and middle (blue) traces were recorded with 5 atm stagnation pressure of helium buffer gas with a laser intensity of $10^5$ and $10^6$ W $cm^{-2}$, respectively, while the bottom (green) trace was recorded for the pyridine neat expansion with the laser intensity of $10^5$ W $cm^{-2}$. The fragment translational energy of the pyridine monomer cation (middle panel) and the power dependence of the pyridine monomer cation signal (right panel) under various conditions. Also shown is the power dependence of the pyridine dimer cation signal. The middle panel also shows quadrant symmetrized velocity-mapped images of pyridine cation corresponding to each of the mass spectra shown in the left panel. The colour coding in all three panels is the same. The mass spectrum of the pyridine neat expansion is in good agreement with that reported in the NIST database. The two signals at 106 and 110 observed in the mass spectrum corresponding to pyridine neat expansion (green trace) could not be assigned.

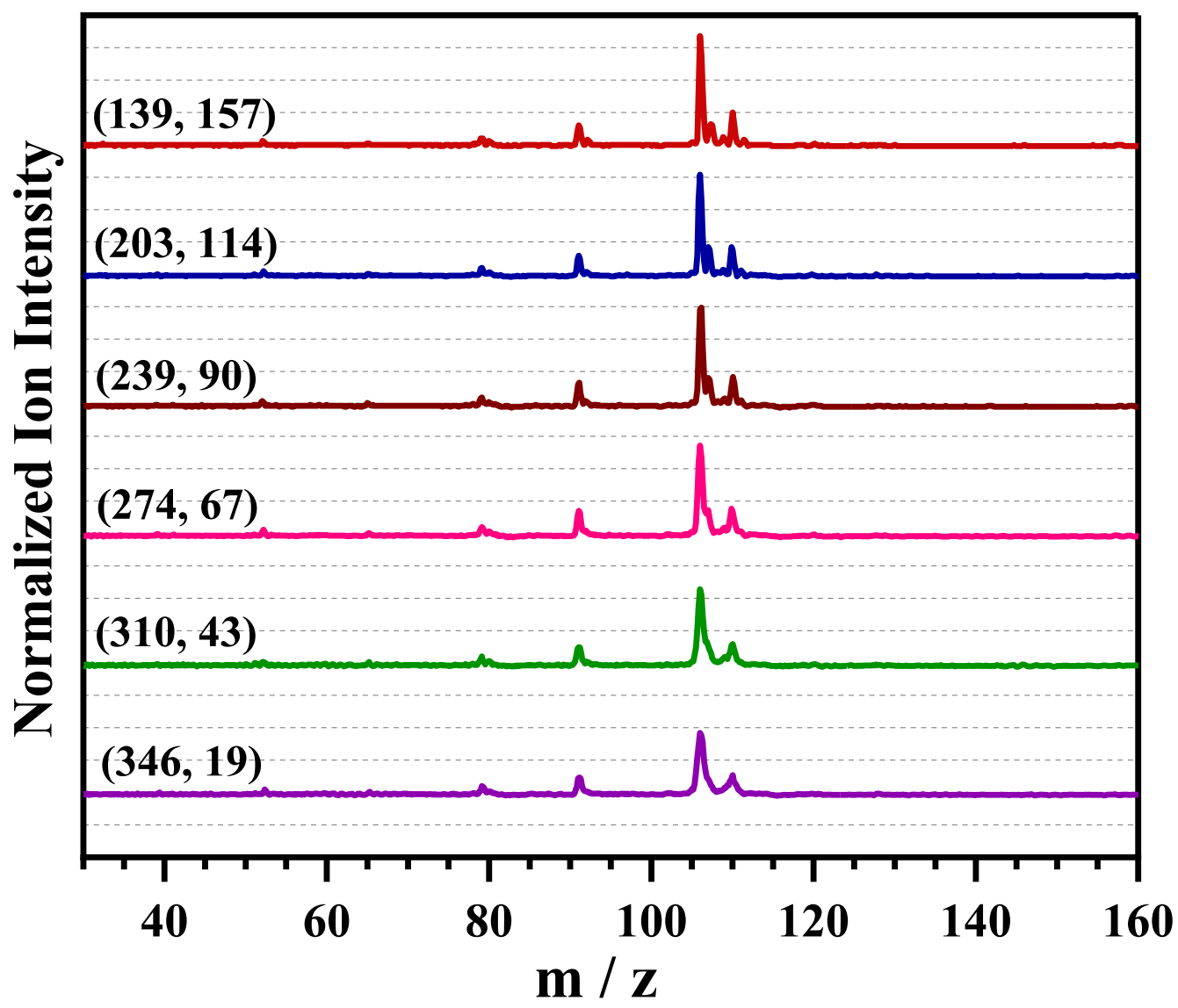

Figure 2. Mass spectra recorded for the neat and unskimmed expansion of pyridine, irradiated by unfocused 266 nm laser with intensity of about $8 \times 10^5$ W cm$^{-2}$, under varying electric fields in the extraction zone. The numbers corresponding to each mass spectrum indicate the electric fields (E1, E2) (in V cm$^{-1}$) between the repeler and lens-1, lens-1 and lens-2, respectively (see Ref 22 for the electrostatic lens arrangement). Each spectrum is normalized with respect to the pyridine signal. The signals at 91, 106 and 110 amu increase from bottom trace to top, which occurs due to varying the electrostatic fields, suggesting that these mass signals originate from ion–molecule collision process.

The broad translational energy distribution of the pyridine cation following a three-photon absorption cannot originate from the pyridine monomer (electron carries all the kinetic energy), therefore pyridine dimer must contribute to the observed translational energy spectrum. A simple energy level scheme is proposed (see Figure 3), wherein an initial two-photo absorption (9.32 eV) leads near-threshold ionization in the Franck-Condon region for the π-stacked anti-parallel displaced pyridine dimer,[13] thereafter absorption of an additional photon (the third photon) results in dissociative ionization of the pyridine dimer. The pyridine cation thus produced will acquire translational energy depending on the nature of the dissociative potential energy curve. The dynamics of the pyridine dimer cation leading to fragmentation were investigated using Born-Oppenheimer molecular dynamics (BOMD) simulations and the trajectories simulated for temperatures 150 K and above resulted in fragmentation with the center-of-mass to center-of-mass (COM–COM) distance greater than 0.7 nm (see Figure 3).[14] The total translational energy distribution profile of the pyridine monomer cation originating from the dissociating dimer cation is in excellent agreement with the experimental observations (see Figure 3). These results substantiate the fact that the pyridine cation originates out of the pyridine dimer cation following a three-photon process. Further, the simulations also reveal that the dissociation of the pyridine dimer cation can either be direct or roaming mediated (see Figure 3).[19] Thus, it can be inferred that pyridine dimer is an exclusive harbinger of the pyridine monomer signal with a laser fluence of around $10^5$ W $cm^{-2}$ (see Figure 1). These results suggest that collective excitation[20] of the pyridine dimer outweighs excitation of the pyridine monomer, similar to observations made in helium nanodroplets.[21]

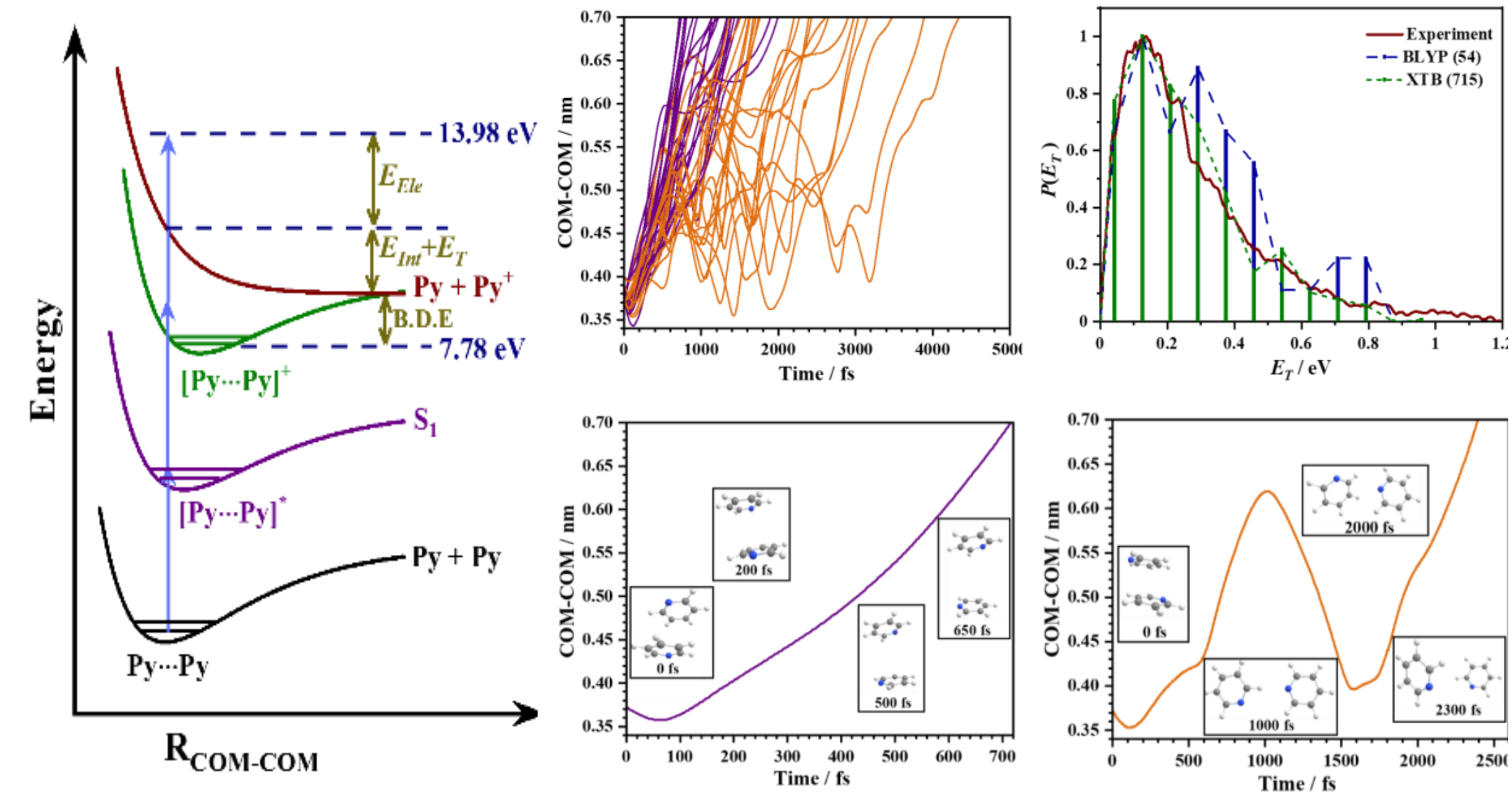

Figure 3. (Left) The schematic diagram (not to scale) for the one-dimensional potential energy curves for the dissociative ionization of pyridine dimer. (Top middle) Center-of-mass to Center-of-mass distance (COM-COM) between the two pyridine moieties plotted as a function of simulation time calculated at BLYP/DZVP level. Of the total of 54 trajectories, some trajectories showed direct dissociation (26 trajectories, purple traces) while the rest showed roaming mediated dissociation (28 trajectories, orange traces). (Top right) Comparison of the total translational energy distribution at two different levels of theory (histogram) with the experimental observation (brown trace). The green and blue dashed lines connect the top of the histogram to show the distribution function. Notice that the comparison between the XTB distribution profile (green curve) and the experimental observation is excellent since a large number of XTB trajectories (715) were calculated. (Bottom middle and right) The snapshots of the dimer cation complex along the simulation time of a single representative trajectory for the direct and roaming mediated dissociation.

The present set of experimental results in combination with the Born-Oppenheimer molecular dynamics simulations corroborate the fact that the three-photon dependence on the appearance of the pyridine cation is due to dissociative ionization of the pyridine dimer and not from the intermolecular Coulombic decay in unbound pyridine monomers as suggested earlier.[1]

## Methods

Pyridine was photoexcited by 266 nm radiation in two different molecular beam configurations; one as neat pyridine expansion and the other by seeding pyridine in helium buffer gas at 5 atm backing pressure. The laser intensities were varied in the range of $10^5$ and $10^6$ W $cm^{-2}$. The resulting pyridine monomer cations were imaged using a velocity map imaging (VMI)spectrometer,[22] and the corresponding translational energy distribution profiles were extracted using the BASEX method.[23] BOMD simulations on the pyridine dimer cations were carried out in an NVE ensemble at 150-200 K with two different density functional theory methodologies viz., XTB (715 trajectories) and BLYP/DZVP (54 trajectories) using CP2K software.[24] For each dissociative trajectory, the translational energy in the centre-of-mass frame was calculated when the COM-COM distance reached 0.7 nm. Details of the methodology are provided as supplementary information.

## Acknowledgments

S.K. and A.T. are supported by PMRF and Institute Postdoctoral fellowships, respectively. This study is based upon the work supported in part by the Science and Engineering Research Board of the Department of Science and Technology (Grant no. CRG/2022/005470) and the Board of Research in Nuclear Sciences (BRNS Grant no. 58/14/18/2020) to GNP. The support and the resources provided by 'PARAM Kamrupa Facility' under the National Supercomputing Mission, Government of India at the Indian Institute of Technology Guwahati are gratefully acknowledged. The authors wish to thank Dr. Rupayan Biswas for his help in computations.

**Author contributions**

S.K. performed all of the experiments; A.T. performed all the calculations. The results were interpreted jointly by all the authors.

**Competing interests**

The authors declare no competing interests

**Data Availability**

The manuscript and the supplementary information contain all the data that has been used to interpret the results.

## SUPPLEMENTARY INFORMATION

## Methods

### Experimental details

The complete details of the experimental set-up can be found elsewhere,[1] Briefly, either neat pyridine or pyridine seeded in helium gas at 5 atm backing pressure, were expanded through a 0.5 mm diameter pulsed nozzle (Series-9; General Valve Corporation) operating at 10 Hz. The free-jet formed in the source chamber was skimmed with a skimmer of 1 mm diameter and introduced into the detection chamber. The ambient pressures of the source and the detection chambers were around $5x10^{-6}$ and $2x10^{-7}$ Torr, respectively. Additionally, experiments with neat and unskimmed pyridine free-jet were also carried out. The pyridine molecules were excited with unfocussed 266 nm radiation using the fourth harmonic of a Nd:YAG laser (Brilliant-B, Quantel) at a 10 Hz repetition rate and the intensities were maintained in the order of $10^5$ and $10^6$ W $cm^{-2}$. The plane of polarisation of the laser was kept parallel to the plane of the detector. The resulting cations were imaged using a four-electrode velocity map imaging (VMI) spectrometer[2] by focussing onto a 50 mm diameter two-stage microchannel plate (MCP) coupled to a P47 phosphor screen (MCP-50DLP47VF; Tectra). The ion cloud was selectively detected by mass gating the front plate of the MCP detector by a 75 ns high voltage pulse (HTS 40-06-OT-75, Behlke). The images were collected using a high-performance, USB, GigE CMOS camera (IDS Imaging Development Systems). The raw images acquired using NuAcq software[3] were quadrant symmetrized using ImageJ software.[4] Abel inversion was carried out by the Basis Set Expansion (BASEX)[5] method to extract the translational energy distribution profile.

### Computational details

The dissociation dynamics of pyridine dimer cation were investigated using Born-Oppenheimer molecular dynamics (BOMD) simulations on CP2K-6.2. software package.[6] The coordinates of the most stable, π-stacked and displaced anti-parallel, structure for pyridine dimer were taken from Ref. 7 7} as the input structure, which was kept at the centre of a cubic box (25 Å) with PBE, and the system was equilibrated by NVT ensemble using a Nose-Hoover thermostat[8] at 30K for a simulation time of 1.4 ps. Starting with several randomly chosen structures from the NVT ensemble, unrestrained BOMD simulations were carried out in an NVE ensemble in the temperature range of 70-200 K with an integration time step of 1.0 fs for a maximum duration of 10 ps. The simulations were carried out using QUICKSTEP module with Becke-Lee-Yang-Parr (BLYP) functional in combination with double ζ valence potential (DZVP) basis set and an auxiliary plane wave was used to expand the valence shell electron density along with Goedecker-Teter-Hutter (GTH) pseudopotentials for the core electrons.[9,10] The electron density was represented using the hybrid Gaussian and plane wave (GPW) method with a cut-off of 400 Ry. Of a total of 129 trajectories, 54 trajectories resulted in fragmentation leading to the formation of pyridine monomer and pyridine monomer cation.[11] Additionally, around 715 NVE trajectories of pyridine dimer cation using XTB[12] methodology at 180 K, starting from randomly chosen structures from the NVT ensemble were simulated. The center-of-mass to Center-of-mass distance (COM-COM) between the two pyridine moieties was monitored as a function of simulation time and deemed fragmented when the COM–COM distance was greater than 0.7 nm and the translational energy in the centre-of-mass frame was calculated.

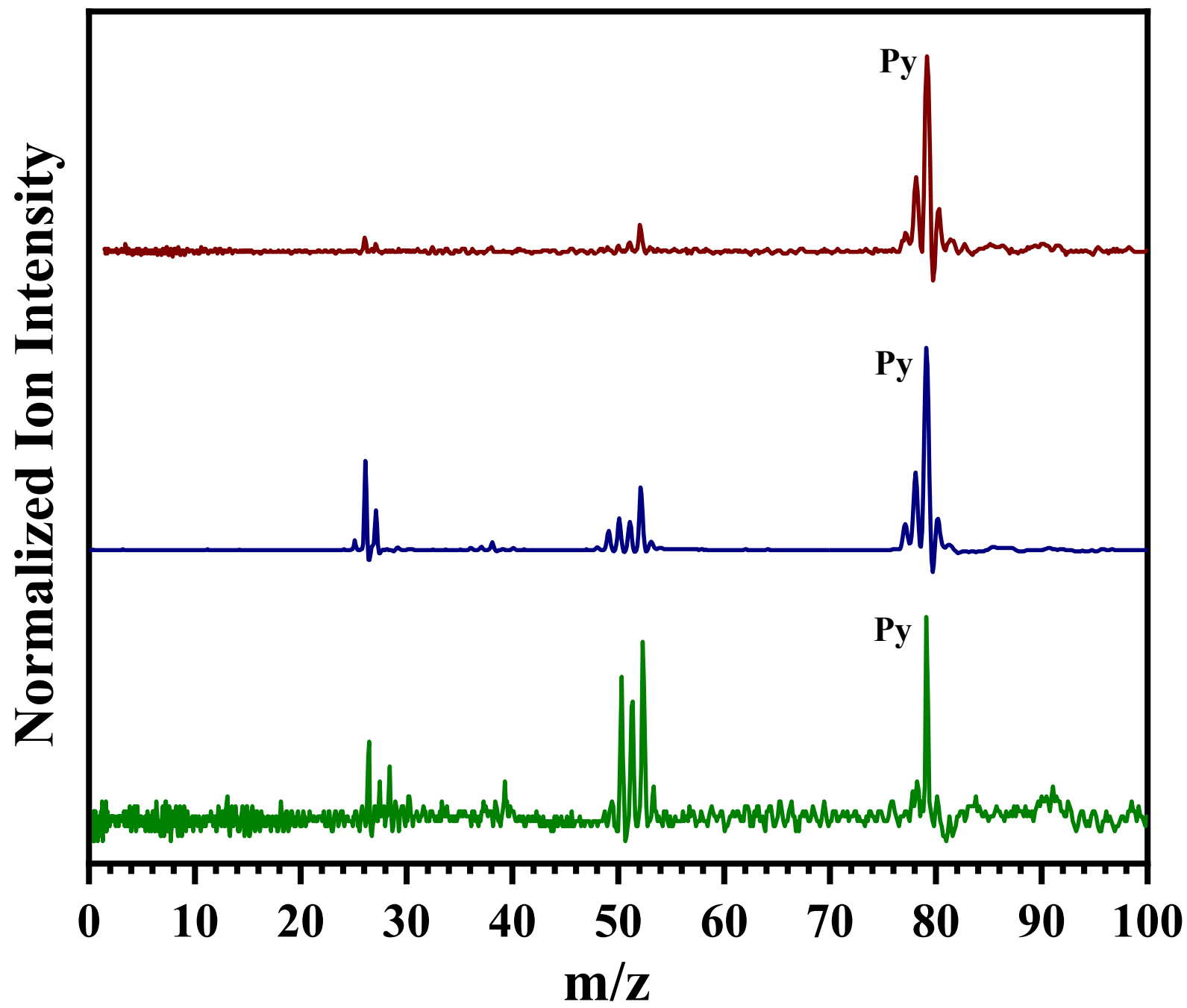

**Figure S1.** Mass spectrum of pyridine recorded under various conditions. The top (brown) and middle (blue) traces were recorded for pyridine seeded in helium buffer gas at 5 atm stagnation pressure with a laser intensity of $10^5$ and $10^6$ W cm$^{-2}$, respectively, while the bottom (green) trace was recorded for the pyridine neat expansion with the laser intensity of $10^5$ W cm$^{-2}$.